\documentclass{article}
\usepackage[T1]{fontenc}
\usepackage[utf8]{inputenc}
\usepackage{ismir}
\usepackage{amsmath,cite}
\usepackage{graphicx}
\usepackage{color}
\usepackage{multirow}
\usepackage{amsfonts}
\usepackage{kotex}
\usepackage{booktabs}
\usepackage{xurl}
\usepackage[bookmarks=false,hidelinks]{hyperref}

\title{Pitch Contour Tokenization using VQ-VAE and Its Application on Korean Traditional Music Analysis}

\multauthor
  {Seonguk Ju$^1$ \hspace{1cm} Seola Cho$^1$ \hspace{1cm} Sooin Chung$^1$}
  {{\bf Danbinaerin Han$^2$ \hspace{1cm} Dasaem Jeong$^1$}\\
  $^1$ Music \& Arts Learning (MALer) Lab, Sogang University, South Korea\\
  $^2$ Graduate School of Culture Technology, KAIST, South Korea\\
  {\tt\small \{jinin5113, dasaemj\}@sogang.ac.kr}}

\def\authorname{S. Ju, S. Cho, S. Chung, D. Han and D. Jeong}

\begin{document}

\maketitle

\begin{abstract}
Computational analysis of music often relies on discrete representations, yet many musical traditions are organized around continuous pitch movement that resists segmentation into note-like units. For such traditions, the discrete units that analysis would build on are not given in advance. We address this gap by learning a vocabulary of local pitch-contour patterns directly from unlabeled audio, using a VQ-VAE that quantizes fixed-length contour segments into a finite codebook. To make the learned tokens stable across segmentation positions and small variations in timing and pitch range, we train the model with a reconstruction objective evaluated under the best alignment among a set of candidate temporal and pitch-domain transformations. Applied to Korean traditional music, the learned tokens recover information about expert-defined \textit{sigimsae} categories without supervision, and in \textit{pansori} individual tokens align with the two principal modes, \textit{Gyemyeonjo} and \textit{Ujo}, supporting their use as units for corpus-level analysis of contour-centric traditions. 
\end{abstract}

\section{Introduction}\label{sec:introduction}

\begin{figure}[t]
  \centering
  \includegraphics[alt={alt.},width=0.9\linewidth]{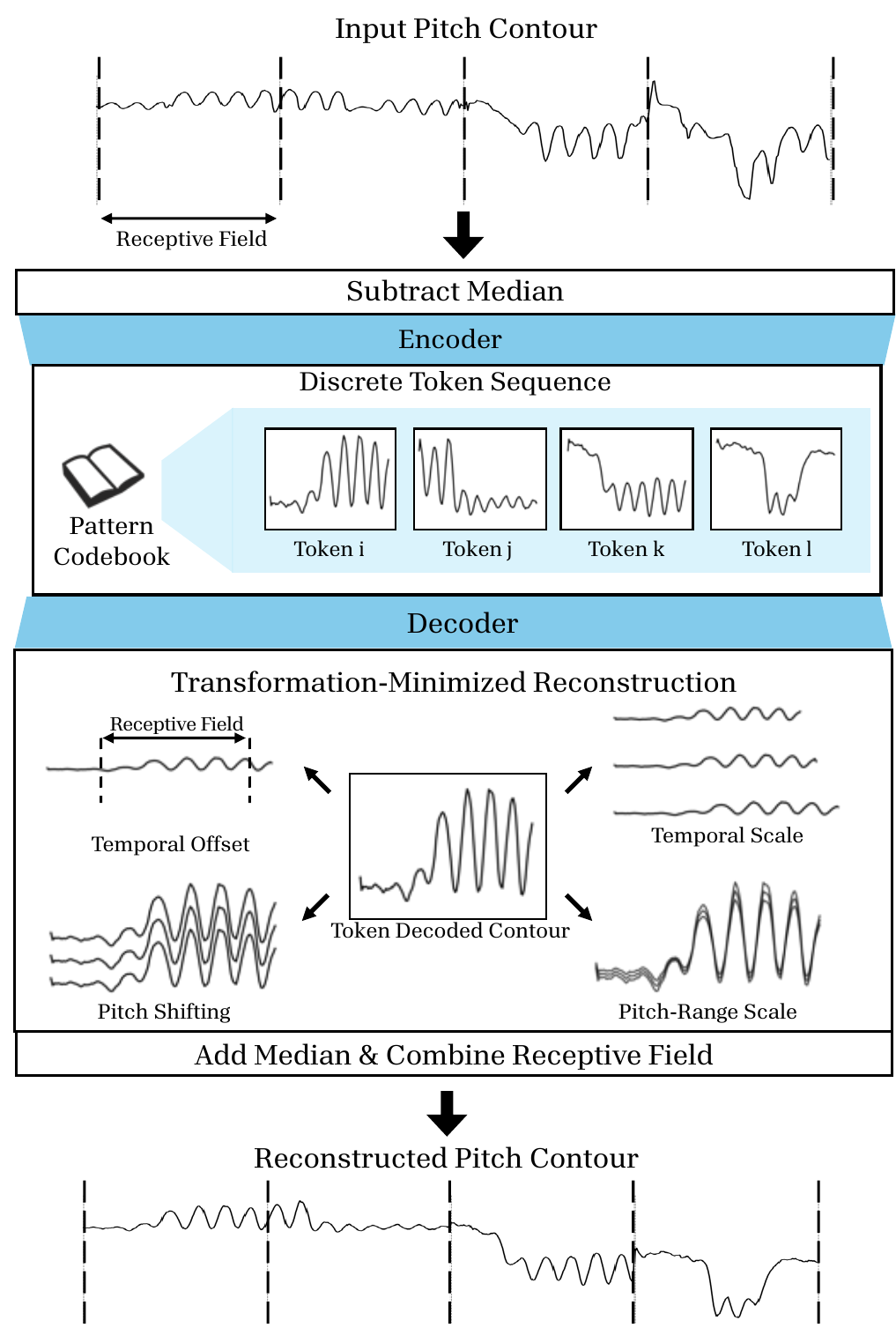}
  \caption{Overview of the proposed model.}
  \label{fig:overview}
\end{figure}

Representing music as sequences of discrete units has been one of the central abstractions underlying computational musicology and symbolic music information retrieval. Its analytical value lies in the fact that transitions between musical events can be counted and modeled statistically. For example, corpus-based chord progression analysis~\cite{moss2019statistical} can construct transition matrices between chord symbols and examine their distributional properties, which is possible because harmonic events are represented as discrete categories. A similar principle underlies probabilistic models of melody: once a melody is represented as a sequence of symbolic events, the probability or information content of each note can be estimated from its musical context. Temperley's theory of uniform information density in music builds on this idea, explaining musical organization in terms of the probability and information density of musical events~\cite{temperley2014information}. The same logic also extends to performance data: MIDI represents performances as note events with onset time, duration, pitch, and velocity, allowing timing and dynamics to be analyzed as event-level attributes~\cite{widmer2003search}. In this sense, symbolic representations are not merely encodings of music, but interfaces that make musical structure available to computational analysis.

The utility of discrete representations, however, is fundamentally contingent on the availability of appropriate discrete units. Automatic music transcription (AMT) has made substantial progress in converting audio into note-level symbolic representations, yet existing systems are developed and evaluated primarily on Western music~\cite{chang2024yourmt3+}. This limitation is not merely a matter of musical style or timbral diversity; it reflects an assumption that musically meaningful events can be segmented into note-like units with relatively well-defined pitch and duration. Many musical traditions do not fully conform to this premise. In such traditions, expressive content is often realized as continuous pitch movement rather than as a succession of clearly bounded note events, making conventional transcription pipelines ill-suited as a basis for analysis.

One way to obtain discrete units in such traditions is to rely on human expertise. 
Some studies use expert-labeled events directly: for instance, a recent study on Carnatic music used expert-labeled svara and svara-form annotations to show that the svara-form performed for a given svara is strongly influenced by neighboring svaras~\cite{DBLP:conf/icassp/NuttallSP25}. 
Others define a vocabulary of intermediate units in advance and transcribe recordings into it: the IDTAP platform, for example, represents Hindustani melodic practice using manually specified pitch-contour trajectories as a middle-level representation between raw F0 curves and conventional notation~\cite{jonathan_myers_2025_17706467}. 
Both directions show that discrete representations of contour-centric music can support analysis, but they share a reliance on predefined categories and manual labor, which is costly to scale and ill-suited to traditions where the relevant ornamental vocabulary is not formally codified. 
This motivates a different question: can we learn a discrete vocabulary of contour units directly from data?

In this work, we investigate this question through VQ-VAE-based pitch contour tokenization. 
VQ-VAE provides a natural framework for this purpose because it learns a finite set of latent codes while preserving enough information to reconstruct the input contour. We use this mechanism not as a generative objective in itself, but as a way to induce a compact vocabulary of local contour patterns that can be used for musical analysis. 
We apply this approach to \textit{sigimsae}, the ornamental system of Korean traditional music, where expressive musical units are often realized as continuous pitch movements rather than conventional note events.

Our contribution is threefold: (i) a VQ-VAE-based pitch-contour tokenization model that learns a discrete vocabulary of local contour patterns without labels; (ii) a transformation-minimized reconstruction loss that makes token learning robust to small temporal and pitch-domain variations; and (iii) an evaluation via segmentation-shift consistency, \textit{sigimsae} classification, and qualitative mode analysis in \textit{pansori}. Our code and a demo page with audio examples are publicly available.\footnote{Codebase: \url{https://github.com/SeongUkJu/pitch-contour-tokenizer}, Demo: \url{https://seongukju.github.io/pitch-contour-tokenizer-page/}}

\section{Related Work}

\subsection{Pitch-Contour Abstractions for Ornamental Music}\label{subsec:ornament}
A notable precedent is found in Indian art music, where quantized pitch contours have served as an intermediate representation capturing ornamental detail beyond note-level transcription.
Ranjani et al.~\cite{DBLP:conf/ismir/RanjaniPS17,10.1121/1.5087277} mapped each pitch value onto a music-theoretically defined grid of note positions and showed that quantizing only at contour critical points preserves r\={a}ga and gamaka characteristics. 
Shikarpur et al.~\cite{DBLP:conf/ismir/ShikarpurDWCH24} adopted finely quantized pitch contours (10-cent bins) as an intermediate representation for hierarchical generative modeling of Hindustani vocal music. 
Both discretize pitch \textit{values} onto a predefined grid; we instead learn a vocabulary over local pitch \textit{patterns} (fixed-length contour segments) directly from unlabeled data.

A separate line of work mines recurring melodic patterns from raw contours via similarity search. 
Gulati et al.~\cite{7081557} used DTW-based similarity over a 365-hour Carnatic corpus, and Nuttall et al.~\cite{DBLP:conf/cmmr/NuttallPPS21} applied the matrix profile to find repeated motifs within a single recording. 
These methods identify a small set of variable-length, recurrence-based instances that highlight reuse, but they do not yield a reusable vocabulary that covers every frame. 
Our tokenizer, by contrast, aims to produce a dense token sequence over the entire contour using a fixed codebook, which makes the resulting representation directly amenable to corpus-level statistical analysis.

\subsection{VQ-VAE and Discrete Representation Learning}\label{subsec:vqvae}
VQ-VAE~\cite{DBLP:conf/nips/OordVK17} learns discrete latent codes through
reconstruction-based training. Given an input $\mathbf{x}$, an encoder
$f_\theta$ produces a continuous embedding $\mathbf{z}_e = f_\theta(\mathbf{x})$,
which is replaced by its nearest entry in a learned codebook
$\mathcal{E}=\{\mathbf{e}_k\}_{k=1}^{K}$ via
$\mathbf{z}_q = \mathbf{e}_{k^\ast}$, $k^\ast = \arg\min_k \|\mathbf{z}_e - \mathbf{e}_k\|_2$,
and a decoder $g_\phi$ reconstructs $\hat{\mathbf{x}} = g_\phi(\mathbf{z}_q)$.
The model is trained with
\begin{equation}
\mathcal{L} = \mathcal{L}_{\mathrm{rec}}
+ \|\mathrm{sg}[\mathbf{z}_e]-\mathbf{z}_q\|_2^2
+ \beta\,\|\mathbf{z}_e-\mathrm{sg}[\mathbf{z}_q]\|_2^2,
\label{eq:vqvae}
\end{equation}
where $\mathrm{sg}[\cdot]$ is the stop-gradient operator and gradients flow
through the quantization step via the straight-through estimator.

In music and audio, vector quantization has been used primarily for generation and compression of audio waveforms or symbolic sequences~\cite{DBLP:conf/ismir/HanILL22, DBLP:journals/corr/abs-2005-00341, DBLP:journals/taslp/ZeghidourLOST22, DBLP:journals/tmlr/DefossezCSA23}, where codes are optimized for fidelity or bitrate rather than interpretability.

Our use of VQ-VAE differs from both lines: the input is a one-dimensional
pitch contour $\mathbf{x}\in\mathbb{R}^{L}$, and the learned codes are
intended as interpretable analytical units rather than for generation or
compression. 
In speech, vector-quantized autoencoders applied to raw or spectral audio have been shown to discover units that align with phonetic content~\cite{DBLP:journals/taslp/ChorowskiWBO19}, supporting the use of VQ-VAE for unsupervised unit discovery from continuous signals.

\section{VQ-VAE on Pitch Contour}\label{sec:page_size}

The overall architecture is based on VQ-VAE, using a stack of 1D convolutional layers, with two key design choices to enable interpretable, segmentation-robust (position-invariant) tokenization of pitch contours, as illustrated in \figref{fig:overview}.

\subsection{Receptive Field Separation}

To ensure a 1:1 correspondence between each discrete token and a fixed-length pitch contour segment, the input is pre-segmented into non-overlapping windows of a fixed size before encoding. 
This prevents inter-token dependency during decoding and promotes interpretability of the learned codebook by constraining each token to represent a single localized contour segment.

To direct the codebook toward capturing ornamental contour patterns rather than absolute pitch levels, we subtract the median pitch value of each receptive field segment before encoding. 
This makes the objective insensitive to absolute pitch offset and encourages each entry to represent local movements such as oscillations and slides. 
When needed for visualization of the full contour sequence, the decoded contour can be shifted back by adding the subtracted median.

\subsection{Transformation-Minimized Reconstruction Loss}
\label{subsec:transrec}
Because the input contour is divided into fixed-size non-overlapping segments before encoding, the assigned token can be sensitive to the arbitrary segmentation position. 
For example, if a vibrato pattern is shifted by only a few frames, the phase observed within a segment changes, although the underlying oscillatory pattern remains the same. 
This matters because the model is trained with a frame-level MSE reconstruction loss: 
even when two contours correspond to the same local ornamental pattern, small temporal misalignments can lead to large reconstruction errors. 
Without an alignment-aware objective, the model may encode incidental frame-level differences as separate entries rather than a stable representation of the underlying pattern.

We address this issue by minimizing the reconstruction error over four transformation factors: \textbf{(i) temporal offset}, which accounts for arbitrary segmentation position; 
\textbf{(ii) temporal scale}, which accounts for small duration differences; 
\textbf{(iii) pitch offset}, which accounts for residual vertical mismatch between the target and reconstruction; and 
\textbf{(iv) pitch-range scale}, which accounts for amplitude or range differences in otherwise similar contour shapes. 
We define a set of candidate transformations consisting of temporal alignment, pitch shifting, and pitch-range scaling. 

Let $\mathbf{x}\in\mathbb{R}^{L}$ be a median-normalized input segment and $\hat{\mathbf{x}}\in\mathbb{R}^{L'}$ be the decoder output, where $L' > L$. 

We define a transformed reconstruction candidate as

\begin{equation}
T_{s,\tau,a,b}(\hat{\mathbf{x}})=a \cdot C_{\tau}\left(R_s(\hat{\mathbf{x}})\right) + b,
\end{equation}
where $R_s(\cdot)$ resamples the decoded contour with temporal scale $s$, $C_{\tau}(\cdot)$ extracts a length-$L$ crop at temporal offset $\tau$, $a$ scales the pitch range, and $b$ shifts the pitch level. 
A length-$L$ crop is then extracted with temporal offset $\tau \in \mathcal{T}$:

\begin{equation}
\hat{\mathbf{x}}^{(s,\tau)} = C_{\tau}\left(\hat{\mathbf{x}}^{(s)}\right),
\end{equation}

where $C_{\tau}(\cdot)$ returns a contiguous segment of length $L$ starting at offset $\tau$. 

The reconstruction loss is computed as the minimum MSE over all candidate transformations:

\begin{equation}
\mathcal{L}_{\mathrm{rec}}
=\min_{s \in \mathcal{S},\ \tau \in \mathcal{T},\ a \in \mathcal{A},\ b \in \mathcal{B}}\frac{1}{L}\left\|\mathbf{x}-\left[a \cdot C_{\tau}\left(R_s(\hat{\mathbf{x}})\right) + b\right]\right\|_2^2 .
\end{equation}

Only the transformation attaining the minimum error is used for backpropagation.

This can be interpreted as fitting the model under the best alignment between the decoded contour and the target segment, rather than penalizing all frame-level discrepancies directly. This objective encourages the codebook to represent contour patterns that are stable under small temporal and pitch-domain variations. 

For example, vibrato contours with similar periodic structure can be mapped to the same token even when they differ in phase or amplitude, because the loss evaluates reconstruction after the best temporal and pitch-domain alignment.

\section{Experimental Setup}\label{sec:experiment}

\subsection{Dataset}
In this study, we apply VQ-VAE to pitch contours of Korean traditional music. Although the proposed method is applicable to any pitch contour-based musical tradition, such as Indian art music or pop vocal singing, we focus on Korean traditional music. 
This is because continuous pitch movement is not merely an expressive embellishment but a central, defining element of the tradition. Unlike Western staff notation, in which music is conceived as a sequence of discrete note events, Korean traditional music, particularly its ornamental system \textit{sigimsae}, is realized through fluid, continuous pitch trajectories in which the boundary between one ``note'' and the next is often ambiguous or undefined. This makes it an especially well-suited testbed for a model designed to discover meaningful structure from pitch contours without presupposing discrete note events.

We selected an in-house \textit{pansori} recording dataset consisting of approximately 280 hours of audio recordings.
\textit{Pansori} is a representative vocal genre of Korean traditional music, sung by a solo singer with barrel drum accompaniment. 
It is well known for its extensive use of melodic ornamentation, making it a particularly rich source of the ornamental patterns investigated in this study.
For \textit{pansori} recordings, the vocal tracks are first isolated using HT-Demucs ~\cite{DBLP:conf/icassp/RouardMD23} to remove instrumental accompaniment.

\subsubsection{Pre-processing}
We extract the fundamental frequency (F0) with the pretrained CREPE~\cite{DBLP:conf/icassp/KimSLB18}, which estimates F0 values every 10 ms with a confidence score in $[0,1]$.
The extracted F0 values are converted to MIDI scale and normalized by median. 
Frames below a predefined confidence threshold are filtered out. 
Within each receptive field segment, the gaps of up to two consecutive filtered frames are filled via linear interpolation. 

Segments containing extensive unvoiced regions are discarded because our 1D CNN-based tokenizer assumes a dense continuous contour and cannot directly represent missing pitch values. 
Filling such regions with a constant padding value would introduce artificial patterns into the input, which could be absorbed by the codebook and hinder the interpretability of the learned representations in subsequent analysis. 
Addressing unvoiced regions more explicitly is left for future work. 

Before filtering, the data are split 8:1:1 into train, validation, and test sets.
During training, a random start point is sampled from each filtered segment at every epoch, introducing variability in segmentation position, while the validation and test sets use fixed segments for consistent evaluation.

\subsection{Model Specification}
The encoder and decoder each consist of 6 stacked one-dimensional convolutional layers with a receptive field size of 128 frames and a codebook of 256 entries. 
The model is trained using the VQ-VAE objective~\cite{DBLP:conf/nips/OordVK17}, in which the standard reconstruction loss is replaced by the transformation-minimized reconstruction loss described in Section~\ref{subsec:transrec}, combined with a vector quantization loss with commitment coefficient $\beta = 2.0$.
Training is conducted for 50K updates using the AdamW optimizer with an initial learning rate of 0.001 and a batch size of 1024. 
We select the checkpoint with the lowest validation reconstruction loss.

The transformation parameters are set empirically. 
The decoder produces $2L$ frames, resampled by $R_s(\cdot)$ to lengths $L_s$ approximately $\mathcal{S}=\{1,\,1.08,\,1.2\}$ times $L$ (i.e., $128$, $139$, and $154$ frames); all valid offsets $\mathcal{T}=\{0,1,\dots,L_s-L\}$ are considered as crops, where $L_s$ is the resampled length. 
Pitch-range scaling and pitch shifting use $\mathcal{A}=\{0.85,1.0,1.15\}$ and $\mathcal{B}=\{-0.1,0.0,0.1\}$.

For the comparison, we also trained an \textbf{Autoencoder Baseline}, adapting Nuttall et al.~\cite{DBLP:conf/dlfm/NuttallSP24}, who applied Uniform Manifold Approximation and Projection (UMAP)~\cite{DBLP:journals/corr/abs-1802-03426} and K-Means clustering to supervised GRU embeddings. 
Our unsupervised variant trains an autoencoder on the \textit{pansori} dataset and tokenizes its latent embeddings via z-score normalization, UMAP, and K-Means.

\section{Segmentation Shift Consistency}
\begin{table}[t]
\centering
\small
\setlength{\tabcolsep}{5.5pt}
\renewcommand{\arraystretch}{1.12}
\begin{tabular}{lcccc}
\toprule
\multirow{2}{*}{Method}
& \multicolumn{2}{c}{KLD $\downarrow$}
& \multicolumn{2}{c}{Acc. $\uparrow$} \\
\cmidrule(lr){2-3}\cmidrule(lr){4-5}
& \scriptsize 0--60
& \scriptsize 0--124
& \scriptsize 0--60
& \scriptsize 0--124 \\
\midrule
Autoencoder
& 1.747 & 2.085 & 0.100 & 0.115 \\
VQ-VAE
& 0.657 & 0.859 & 0.462 & 0.488 \\
\quad + temporal align
& \textbf{0.531} & \textbf{0.735} & \textbf{0.519} & \textbf{0.561} \\
\quad + all transformations
& 0.652 & 0.881 & 0.456 & 0.495 \\
\bottomrule
\end{tabular}
\caption{Token consistency evaluation results. KLD denotes Token Distribution KLD, and Acc. denotes Token Matching Accuracy. Results are reported for two segmentation-offset ranges: 0--60 and 0--124 frames. ``+ temporal align'' applies temporal offset and temporal-scale transformations, while ``+ all transformations'' additionally includes pitch offset and pitch-range scaling.}
\label{tab:token_consistency}
\end{table}

We introduced the transformation-minimized reconstruction loss to reduce the sensitivity of token assignment to temporal and pitch-domain variations. 
Here, we isolate one of its intended effects: robustness to segmentation position. 
Specifically, we test whether the same long pitch contour yields similar token sequences when the starting position of the fixed-size segmentation window is shifted.

\subsection{Experimental Dataset Construction}
From the \textit{pansori} dataset used for VQ-VAE training, we select contour segments that satisfy the filtering criteria, contain no unvoiced frames, and are at least three times longer than the receptive field. For each segment, we generate multiple token sequences by shifting the segmentation start position in increments of 4 frames.

We report results for two offset ranges: 0--60 frames and 0--124 frames. The former evaluates shifts smaller than half of the 128-frame receptive field, while the latter covers nearly all possible phases within one receptive field and thus provides a stricter test of segmentation robustness.

\subsection{Evaluation Metrics}

We use two complementary metrics. \textbf{Token Distribution KLD} measures whether the overall codebook usage remains stable across shifted versions of the same contour. For each segment, we construct a reference token distribution by aggregating token counts across all shifted versions. We then compute the Kullback--Leibler divergence (KLD) between the token distribution of each shifted version and this reference distribution, and average the result over versions and segments. Lower KLD indicates more stable token usage.

\textbf{Token Matching Accuracy} measures local token-level consistency. We use the offset-0 token sequence as the reference and compare each shifted token sequence against it. For offsets below half of the receptive field, tokens are compared at the same sequence index. For offsets of 64 frames or greater, each shifted token is compared with the succeeding reference token, since the shifted receptive field overlaps more strongly with the next reference window. Higher accuracy indicates more consistent token assignment across shifts.

\subsection{Results}
As shown in \tabref{tab:token_consistency}, the autoencoder baseline is highly sensitive to segmentation shifts, whereas VQ-VAE substantially improves both distribution-level and local token consistency. 

Adding temporal alignment gives the best results on this segmentation-specific evaluation, supporting the intended effect of the transformation-minimized reconstruction loss. The full transformation setting performs worse than temporal alignment alone on this metric. 
This is expected because the additional pitch-domain transformations are not designed specifically for segmentation robustness; rather, they allow similar contour shapes to be matched after pitch shifting or pitch-range scaling. 
As a result, similar contours may be represented by different tokens whose decoded outputs can still be adjusted to fit the same target, which can reduce exact token consistency under shifted segmentations. 
Their benefit is therefore evaluated more directly in the downstream \textit{sigimsae} classification experiment.

\section{Sigimsae Classification}

If the learned tokens are to serve as analytical units, their patterns should align with expert-defined ornament categories. We test this with a minimal probing setup, a per-token MAP classifier — that isolates how much label-discriminative information is carried by individual tokens.

\subsection{Sigimsae Labeled Dataset}
We employed the annotated Korean traditional music recordings from the NIA AI-Hub dataset~\cite{NIAGugak2022}.
The dataset consists of annotated \textit{sigimsae} segments from monophonic Korean traditional music recordings. 
Because it is highly class-imbalanced and contains both vocal and instrumental performances, we report results in two settings: the full setting, with 34.0 hours of audio and 42.3k annotated segments totaling 9.3 hours across six classes, and the vocal-only setting, with 10.0 hours of audio and 13.5k annotated segments totaling 2.4 hours across five classes.

The data are split at the track level with per-class stratification into train, validation, and test sets (approximately 8:1:1), ensuring that segments from the same track do not appear in multiple splits.

\subsection{Token-to-label lookup classifier}
For each annotated \textit{sigimsae} segment, the center portion is cropped to the receptive field size and encoded into a single token. 
Segments shorter than the receptive field are padded with surrounding context.

We use a simple frequency-based classifier to test how much label-discriminative information is carried by individual tokens. This is not intended to build a strong supervised classifier. Rather, it serves as a probing task: if the tokens learned without \textit{sigimsae} labels are musically meaningful, their distribution should contain information that is predictive of expert-defined ornament categories.

From the training set, we count how often each token occurs with each \textit{sigimsae} label and use these counts to estimate $P(\text{token} \mid \text{label})$. 
Classification is then performed using maximum a posteriori (MAP) estimation:
\begin{equation}
\hat{y} = \arg\max \left[\tau \cdot \log P(y) + \log P(\text{token}\mid y) \right]
\end{equation}
where the temperature $\tau \in [0, 1]$ is selected on the validation set and applied to the test set. 

As a scale reference, we additionally report a supervised Encoder-Decoder Temporal Convolutional Network (ED-TCN), which was employed for R\={a}ga ornamentation detection~\cite{DBLP:journals/corr/abs-2505-04419} trained directly on \textit{sigimsae} classification with median-normalized F0 contours. 
It uses full label supervision and is not directly comparable, but indicates the order of magnitude reachable with supervised contour-based classification.

\subsection{Results}

\begin{table}[t]
\centering
\small
\setlength{\tabcolsep}{5pt}
\renewcommand{\arraystretch}{1.12}
\begin{tabular}{lcccc}
\toprule
\multirow{2}{*}{Method}
& \multicolumn{2}{c}{$F1_{macro}$ $\uparrow$}
& \multicolumn{2}{c}{$mAP$ $\uparrow$} \\
\cmidrule(lr){2-3}\cmidrule(lr){4-5}
& \scriptsize All
& \scriptsize Vocal
& \scriptsize All
& \scriptsize Vocal \\
\midrule
Autoencoder
& 0.284& 0.324& 0.265& 0.290\\
VQ-VAE
& 0.266& 0.319& 0.248& 0.302\\
\quad + temporal align
& 0.285& 0.320& 0.272& 0.317\\
\quad + all transformations
& \textbf{0.285}& \textbf{0.331}& \textbf{0.278}& \textbf{0.324}\\
\hline
ED-TCN~\cite{DBLP:journals/corr/abs-2505-04419}
& 0.463 & 0.541 & 0.579 & 0.531 \\ 
\bottomrule
\end{tabular}
\caption{\textit{Sigimsae} Classification results on NIA AI-Hub dataset~\cite{NIAGugak2022}. We report $F1_{macro}$ and $mAP$ for both all/vocal subset settings.}
\label{tab:sigimsae_classification}
\end{table}

As presented in Table~\ref{tab:sigimsae_classification}, vanilla VQ-VAE alone does not outperform the autoencoder baseline in $F1$. Without alignment-aware training, similar contours are split across distinct codes, fragmenting per-class distributions. Adding the full transformation-minimized loss reverses this and yields the best probing scores across both settings, with a larger gain than temporal alignment alone. 
The gap to the supervised ED-TCN remains substantial, but a single token recovers roughly three-fifths of its $F1$ on the vocal setting without sequence modeling or label supervision during representation learning.

\begin{figure}[t]
  \centering
  \includegraphics[alt={alt.},width=0.9\linewidth]{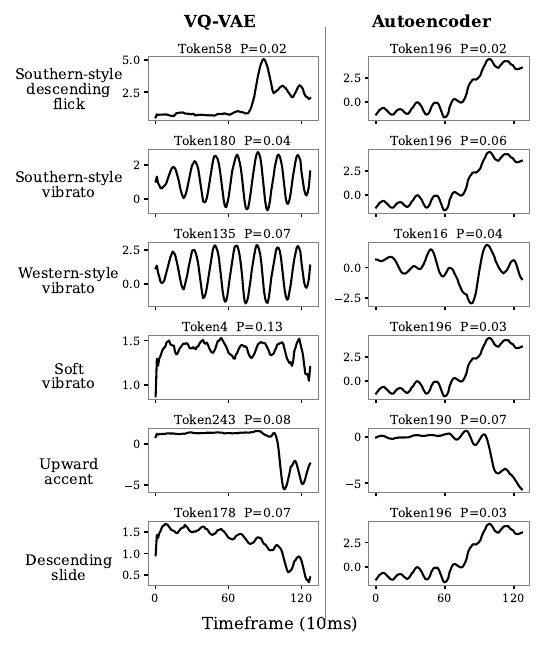}
  \caption{Decoded contours of the most representative token per \textit{sigimsae} category, selected by the highest $P(\textit{token}\mid\textit{label})$, for VQ-VAE (left) and Autoencoder baseline (right).}
  \label{fig:nia_label_token}
\end{figure}

\figref{fig:nia_label_token} visualizes the per-class prototype token (highest $P(\text{token}\mid\text{label}))$  for each model. For VQ-VAE, the six classes recruit six distinct tokens, each visually matching the canonical realization of its category: the upward-leap-and-glide of \textit{southern-style descending flick} (Token 58), the monotonic descent of \textit{descending slide} (Token 178), and three vibrato classes mapped to tokens with clearly distinct oscillation profiles. 
For the autoencoder, a single token (Token 196) is selected as the prototype for four of the six classes. 
The autoencoder's embeddings entangle phase, pitch level, range, and shape, so K-Means cuts across ornament types.
The transformation-minimized loss removes these factors during training, leaving contour shape as the main basis for the codebook.

\begin{figure}[t]
  \centering
  \includegraphics[width=1.0\columnwidth]{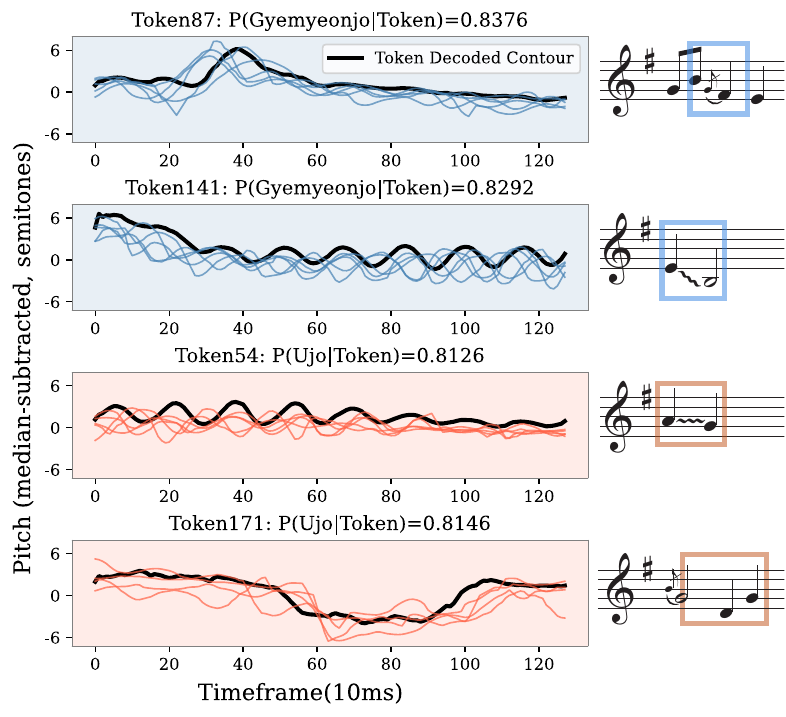}
  \caption{Decoded contours(black) of representative tokens for \textit{Gyemyeonjo} (top) and \textit{Ujo}, selected by the highest $P(\text{mode}\mid\text{token})$. Blue and red lines show the median-subtracted original contours assigned to each token for \textit{Gyemyeonjo} and \textit{Ujo}, respectively.}
  \label{fig:mode_token}
\end{figure}

\section{Token Analysis in Pansori}
As a case study, we explore whether the learned tokenizer can serve as a tool for data-driven musicological analysis by examining how individual tokens relate to the two principal modes of \textit{pansori}.

\textbf{Dataset:}
\textit{Pansori} has two representative modes: \textit{Gyemyeonjo} and \textit{Ujo}, each characterized by distinct melodic and ornamental patterns. 
We used a mode-annotated subset~\cite{anonymous2026pansorimode} of our in-house \textit{pansori} dataset, which was annotated by an expert. The dataset consists of 26.7 hours of \textit{Gyemyeonjo} annotations and 8.7 hours of \textit{Ujo} annotations. 

\textbf{Method:} A \textit{pansori} mode is largely defined by its scale, but each mode is also known to carry characteristic ornamental patterns. Because our tokens are median-normalized and thus carry no absolute pitch information, any mode-specific token usage must arise from contour shape itself. We therefore count token occurrences within mode-labeled segments and identify the tokens with the highest $P(\text{mode} \mid \text{token})$ for each mode, treating these as candidate ornamental patterns characteristic of each mode.

\textbf{Result}: Figure~\ref{fig:mode_token} shows the contour of those representative tokens together with examples of input contours assigned to each token. To provide musical context, one of the co-authors also transcribed a slightly extended window around each example into staff notation.

Inspecting these tokens confirms that each is indeed characteristic of its associated mode. Token 87 captures the upper turn gesture, and Token 141 captures a downward gliding contour; both are widely described in \textit{pansori} studies as defining ornamental features of \textit{Gyemyeonjo} (Score Excerpts 24 and 3 in ~\cite{jo_pansori_kyemyunjo}). 
Tokens 171, in contrast, corresponds to melodic phrase frequently observed in \textit{ujo}.

Token 54 is particularly informative. The descending major-second motion to the tonic that it captures is a melodic move shared by both \textit{Gyemyeonjo} and \textit{Ujo}, yet this specific token occurs predominantly in \textit{Ujo}. The distinguishing factor is the wide vibrato applied to the upper note in \textit{Ujo} before it resolves to the tonic. The tokenizer, trained without any mode supervision, separates the two modes along this ornamental dimension, a distinction that note-level transcription would discard since the underlying pitch sequence is identical.

Although limited to a single case, this result indicates that the learned tokens can serve as analytical units linking raw contours to higher-level musicological categories. The same codebook, queried against other label sets such as schools or performers, offers a path toward corpus-level study of ornamental practice without requiring predefined ornament categories or note-level transcription.

\section{Conclusion and Future Work}
We proposed a VQ-VAE-based tokenizer that learns a discrete vocabulary of local pitch-contour patterns without ornament labels, together with a transformation-minimized reconstruction loss that makes the learned tokens robust to small temporal and pitch-domain variations.
Experiments on Korean traditional music show improved segmentation-shift consistency and \textit{sigimsae} classification over baselines, and qualitative analysis on \textit{pansori} shows that individual tokens separate \textit{Gyemyeonjo} and \textit{Ujo} even when the underlying pitch sequence is shared, a distinction that note-level transcription would discard.

Our approach has one main limitation: our 1D-CNN tokenizer assumes a dense continuous contour and cannot directly model unvoiced regions, so we restricted training and evaluation to segments with few unvoiced frames. Extending the model to handle unvoiced regions explicitly is necessary for tokenizing entire recordings end-to-end.

Two directions remain open for future work. The most immediate is corpus-level musicological analysis: converting long recordings into token sequences and studying transition statistics, n-gram distributions, or conditional patterns of ornamental units across modes, schools, and performers — analyses that have been routine for symbolic Western music but are not yet available for contour-centric traditions. The same framework is also in principle applicable to other such traditions, including Indian art music and pop vocal singing.

\section{Acknowledgement}
This work was supported by the National Research Foundation of Korea (NRF) grant funded by the Korea government (MSIT) (RS-2025-00560548).

\bibliography{ISMIRtemplate}

\end{document}